\documentclass[preprint,12pt,authoryear]{elsarticle}

\usepackage{amsmath,amssymb,amsfonts,amsthm,latexsym,color,url}
\usepackage{multirow}
\usepackage{tabularx}
\usepackage[normalem]{ulem}

\def\sN{{\mathcal N}}

\def\sB{{\mathcal B}}

\def\sA{{\mathcal A}}

\def\sY{{\mathcal Y}}

\def\vectorfontone{\bf}
\def\vectorfonttwo{\boldsymbol}
\def\vx{{\vectorfontone x}}                      
\def\vy{{\vectorfontone y}}                      

\def\vone{{\vectorfontone 1}}

\def\vgamma{{\vectorfonttwo \gamma}}             %
\def\vpi{{\vectorfonttwo \pi}}                   %
\def\vomega{{\vectorfonttwo \omega}}             %

\def\bE{{\mathbb E}}                             
\def\bP{{\mathbb P}}                             
\def\bR{{\mathbb R}}                             

\def\expit{\text{expit}}

\newcommand{\indep}{\raisebox{0.05em}{\rotatebox[origin=c]{90}{$\models$}}}

\newcommand{\wyc}[1]{{\color{red}#1}}

\journal{Computational Statistics and Data Analysis}

\begin{document}

\begin{frontmatter}



\title{Variational Nonparametric Discriminant Analysis}
\author[1]{Weichang Yu\corref{cor1}}
\ead{w.yu@maths.usyd.edu.au}
\author[1]{Lamiae Azizi}
\author[1,2]{John T. Ormerod}
\address[1]{School of Mathematics and Statistics, University of Sydney}
\address[2]{ARC Centre of Excellence for Mathematical and Statistical Frontiers, \\ The
	University of Melbourne, Parkville VIC 3010, Australia}
\cortext[cor1]{Corresponding author's address: Carslaw Building F07, Eastern Ave, Camperdown NSW 2006, Australia.}

\begin{abstract}
Variable selection and classification are common objectives in the analysis of high-dimensional data.
Most such methods make distributional assumptions that may not be compatible with the diverse families of 
distributions data can take. 
A novel Bayesian nonparametric discriminant analysis model that performs both 
variable selection and classification within a seamless framework is proposed. P{\'o}lya tree priors are assigned 
to the unknown group-conditional distributions to account for their uncertainty, and allow prior beliefs about 
the distributions to be incorporated simply as hyperparameters.
The adoption of collapsed variational Bayes inference in combination with a chain of functional approximations
led to an algorithm with low computational cost.
The resultant decision rules carry heuristic interpretations and are related to an existing two-sample 
Bayesian nonparametric hypothesis test. By an application to some simulated and publicly available real 
datasets, the proposed method exhibits good performance when compared to current state-of-the-art approaches.

\end{abstract}

\begin{keyword}
Variational inference \sep 
Bayesian nonparametrics \sep 
P{\'o}lya trees \sep 
Classification \sep 
Variable selection \sep 
High dimensional statistics.

\end{keyword}

\end{frontmatter}


\section{Introduction}
\label{intro}

Discriminant analysis (DA) is a classifier   that has seen several recent applications in 
high dimensional data analysis.
In the context of two response groups (group 1 and group 0), DA classifies a new observation to 
group 1 if the likelihood of the variables and response evaluated at group 1 is greater than 
the  likelihood evaluated at group 0. 
To facilitate the computation of the likelihood, the \emph{group-conditional distribution} 
(the conditional distribution given the response) of the predictor-variables is commonly assumed 
to be Gaussian as with the popular variants such as linear discriminant analysis and quadratic 
discriminant analysis \citep{Fisher1936, McLachlan1992}.

Drawbacks of traditional formulations of DA for high-dimensional data analysis include the lack of variable 
selection options, and restrictive distributional assumptions. 
Variable selection techniques are important because they overcome the gradual accumulation of  
estimation errors as the number of  variables increases \citep{FanLv2010}. 
In addition, erroneous assumptions imposed on the diverse range of distributional families from which 
the variables could be drawn may lead to an inflation of classification errors.
While extensions that perform variable selection are available 
\citep[see for example:][]{Friedman1989, Ahdesmaki, WittenTibs, Yu}, most of these rely on normality 
assumptions that do not hold in general. Other extensions perform 
dimension reduction at the pre-classification stage that sacrifice the interpretability of results 
which is usually required to shed light on underlying  scientific questions 
\citep[see for example][]{Sugiyama2007}.

While solutions such as monotonic transformations \citep[see for example][]{BoxCox, Benjamini2012}
and finite mixture modelling for the group-conditional distributions \citep{Hastie1996, Celeux2010} 
can improve the fit of the data, problems can still arise. In particular, finite mixture modelling
is still subject to model misspecification if the number of mixture components and/or the mixture 
densities are incorrectly specified \citep{Fraley2002}.

Model specification problems can be mitigated by adopting nonparametric approaches as they do not confine the model 
space to a particular parametric family of distributions. An example of the nonparametric approach is 
to estimate the unknown group-conditional distributions with kernel density estimators 
\citep[see][]{HallWand1988, Ghosh2004, Ghosh2006}. However, the density may be undersmoothed in regions 
of the domain space where there are few observations. 

In Bayesian nonparametrics, the unknown 
distributions are regarded as random measures, and assigned a prior on the space of 
possible distributions. A popular option for this prior is the \emph{Dirichlet process mixture} that has 
been incorporated into several DA models \citep[see for example][]{FuentesGarcia2010,Ouyang2017}. One 
alternative has been proposed by \cite{Gutierrez2014}. In their two-stage variable selection and 
classification procedure for high dimensional data, they assigned a Gaussian process prior in a penalised spline 
model to identify informative variables, and then a geometric-weighted mixture model for the classification 
step. 

An alternative nonparametric prior is the P\'{o}lya tree \citep{Mauldin1992}. Unlike the other Bayesian 
nonparametric priors, this prior allows information about the unknown distribution to be incorporated 
as the prior expectation. The P\'{o}lya tree has been applied extensively to density estimation, and Bayesian
nonparametric hypothesis testing problems \citep{Hanson2002, Berger2001, Ma2011, Holmes2015, Cipolli2016, 
Filippi2017}. In particular, \cite{ChenHanson} proposed a Bayesian nonparametric analogue of the multiple 
samples test that demonstrated superior results to a similar model \citep{Holmes2015} by assigning a 
P\'{o}lya tree to each of the unknown population distribution. \cite{Cipolli2016} proposed a 
multiple hypothesis testing framework, and assigned a P\'{o}lya tree hyper-prior for the unknown mean 
parameters of a Gaussian model. However, according to our knowledge, only \cite{Cipolli2018} attempted 
to use multivariate P{\'o}lya tree prior on the joint distribution of the variables in the context of 
classification. Although their proposed method seems to have good performance in most suitable datasets, 
additional pre-processing is required in high dimensional settings.

In this paper, we propose a novel dual-objective Bayesian nonparametric DA model that makes inference for 
variable selection and classification in a unified framework. This is achieved by introducing a set of 
variable selection parameters into the hierarchical framework of the Bayesian DA model, and an independence 
assumption between variables to handle high dimensionality. We assign a P\'{o}lya tree prior to the unknown 
group-conditionals as a representation of our uncertainty and use collapsed variational Bayes inference to 
obtain posterior estimates. This combination of prior choice and posterior inference method lead to a classification 
rule that carries a heuristic interpretation. The heavy computational burden that comes with fitting a 
nonparametric model is reduced to an acceptable range through several computational shortcuts that follow 
from functional approximations. This makes our proposed model appealing for analysing high dimensional data.

Our approach builds on the previous works described above, but differs in several ways related to the use of the 
P\'{o}lya tree prior. In \cite{Cipolli2016}, the P\'{o}lya tree is assigned as a hyperprior to unknown 
Gaussian means of the data, whereas in our model the P\'{o}lya tree is assigned as the prior of the unknown 
group-conditional distributions that the variables are drawn from. \cite{Cipolli2018} assigned a P\'{o}lya 
tree prior on the joint group-conditional distribution of the variables instead of having $p$ univariate 
P\'{o}lya tree priors, as is the case with our model. Both of these papers have been proposed for low 
dimensional data analysis in contrast to our approach that was specifically designed to handle situations in 
which the number of variables ($p$) is greater than the sample size ($n$). Furthermore, we have simplified the 
model for high dimensional datasets by assuming independence between variables and assigned a separate 
P\'{o}lya tree prior to the group-conditional distribution of each variable.

In Section \ref{Model}, we will provide a description of our proposed unified model for high dimensional 
classification and justify the choice of priors. This will also include a brief description of the 
P\'{o}lya tree construction scheme. In Section \ref{Inference}, we will elaborate on the posterior 
inference of our model, and the heuristic interpretation of our resultant classification rule. This will 
be followed by a short Section \ref{ChooseC} that discusses the setting of a hyperparameter in our model. 
In Section \ref{Numerical}, we compare our proposed model with existing options in simulated, and gene 
expression datasets. Circumstances that lead to good performance of the model will also be discussed. 
Finally, we will suggest possible extensions, and conclude in Section \ref{Conclusion}.

\section{Discriminant analysis with variable selection}
\label{Model}

Consider a data set consisting of $n$ observations $\{ (y_i, \vx_i ), i=1,\ldots,n \}$, where 
$\vx_i = (x_{i1}, \ldots, x_{ip})^T \in \bR^p$ are   $p$ predictor variables and $y_i$ is the binary response 
of the $i-$th observation respectively. In this paper, we focus on the case of continuous variables, and 
will defer the discussion of an extension to other variable types to Section \ref{Conclusion}. By 
conditioning on the responses $(y_1, \ldots, y_n)^T$, we assume independence between observations, i.e., 
$\vx_u \, \indep \, \vx_v$ where $u \neq v$, and between variables, i.e., $x_{ih} \, \indep \,  x_{ig}$ 
where $h \neq g$. The latter is a common assumption for high-dimensional DA models known as the 
\emph{na{\"i}ve Bayes} assumption \citep[see][]{FanFan, Ahdesmaki, Witten2011, WittenTibs}. 
This assumption allows us to circumvent  infeasible computations imposed in high dimensional settings ($p \gg n$),
and has been shown to perform  reasonably well under some conditions \citep{Bickel, Yu}.

Here, we describe a modification of the usual 
DA model described in \cite{McLachlan1992} that allows for variable selection 
in the context of pairwise independent 
variables. Given two distributions $F_{j1}$ and $F_{j0}$, the group-conditional distributions of the usual 
discriminant analysis are
$$
x_{ij} \; | \; y_i \stackrel{\mbox{\scriptsize iid.}}{\sim} \left\{ \begin{array}{ll}
F_{j1}, & \mbox{ if $y_i = 1$; and} \\ [1ex]
F_{j0}, & \mbox{ if $y_i = 0$}.
\end{array} \right.
\\ [3ex]
$$
To extend this model to select discriminative variables, we shall introduce a set of binary variable selection parameters
$\vgamma = (\gamma_1, \ldots, \gamma_p)^T.$
Given three distributions $F_j$, $F_{j1}$ and $F_{j0}$, each 
$\gamma_j$ controls the sampling scheme of $X_{ij}$ for $1 \le i \le n$ as follows:
\begin{equation}
\label{ProposedModel1}
\text{If } \gamma_j = 1, \text{ then } \; x_{ij} \; | \; y_i \stackrel{\mbox{\scriptsize iid.}}{\sim} \left\{ \begin{array}{ll}
F_{j1}, & \mbox{if $y_i = 1$;} \\ [2ex]
F_{j0}, & \mbox{if $y_i = 0$;}
\end{array} \right.
\end{equation}

\noindent and if $\gamma_j = 0$, then
\begin{equation} 
\label{ProposedModel2}
x_{ij} \stackrel{\mbox{\scriptsize iid.}}{\sim} F_j.
\end{equation}

\noindent 
Note that $\gamma_j=1$ corresponds to a case in which variable $j$ is discriminative while $\gamma_j= 0$ corresponds to 
a non-discriminative case. 

The binary responses are
distributed as
\begin{align*}
&y_i \; | \; \rho_{y} \stackrel{\mbox{\scriptsize iid.}}{\sim} \text{Bernoulli} (\rho_y),
\end{align*}
where the parameter $\rho_y$ may be interpreted as the prior probability of sampling
an observation from response group 1.

\subsection{Priors for $\rho_y$ and $\vgamma$ }
\label{Priors}
In most applications, the population proportion $\rho_y$ is unknown and 
this has also been the case with the data which we have analysed in Section \ref{Numerical}.
A Bayesian solution is to assign
a hyper prior distribution for $\rho_y$, and a natural choice of prior would be the beta distribution, i.e.,
\begin{equation*}
\rho_y \sim \text{Beta} (a_y, b_y).
\end{equation*}

\noindent 
Due to the beta-binomial conjugacy, this choice leads to a closed form expression
for the joint density of the model marginalised over $\rho_y$ which is useful for posterior inference.

A natural choice of prior for the binary variable selection parameters, $\gamma_1,\ldots,\gamma_p$,
is   the Bernoulli distribution, i.e.,
\begin{align*}
&\gamma_j \; | \; \rho_{\gamma} \stackrel{\mbox{\scriptsize iid.}}{\sim} \text{Bernoulli} (\rho_\gamma), \quad 1\le j\le p.
\end{align*}
The parameter $\rho_\gamma$ may be interpreted as the proportion of discriminative 
variables in the dataset. Following the class of complexity priors \citep{Castillo2015}
that has demonstrated the ability to down-weight 
high-dimension models while allocating sufficient prior 
probability to the true model in several problems, we have chosen the hyper prior
\begin{align*}
\rho_{\gamma} \sim \text{Beta} (1, p^u), \;\; \text{for some } u>1.
\end{align*}
The choice of the constant $u$ affects the penalty of the resultant variable selection rule as described later
in Section \ref{Inference}.

\subsection{Priors for unknown distributions}
\label{PolyaPrior}

\cite{Yu} modelled the distributions $F_{j1}$, $F_{j0}$ and $F_{j}$, $1 \le j \le p$ described
in (\ref{ProposedModel1}) and (\ref{ProposedModel2}) as Gaussian. As discussed in the introduction this is restrictive. Instead we will 
treat the $F$'s in this paper as unknown probability measures 
and assign them a \emph{P\'{o}lya tree prior} \citep{Lavine1992}.
This prior is a distribution defined on a family of distributions on a domain $B$. Hence, one draw from a P\'{o}lya tree
is a particular probability distribution. The process to draw
a distribution  from a P\'{o}lya tree is described in Table \ref{PolyaConstruction}.

\bgroup
\def\arraystretch{1.4}
\begin{table}[ht!] \refstepcounter{table}\label{PolyaConstruction}
	\centering
	\begin{tabular}{l}
		\normalsize
		\bf{Table 1}~ \normalfont{P{\'o}lya tree construction scheme }\\
		\hline
		\normalsize 1. Construct the dyadic tree in Figure \ref{fig::PTree} by specifying recursive partitions\\[-0.08in]  
		of the domain $B$. Note that each tree \emph{layer} is a partition of $B$ and the\\[-0.08in] 
		recursive relationship between successive layers is: $B = B_0 \cup B_1$,  \\[-0.08in] 
		$B_0 = B_{00} \cup B_{01}$, $B_1 = B_{10} \cup B_{11}$ and so on. \\[0.1in] 
		
		\normalsize 2. Set $\Pi = \{B, B_0, B_1, B_{00}, B_{01}, \ldots \}$ as the collection of \emph{partition-subsets}\\[-0.08in] 
		obtained by taking the union of the partitions in step 1.\\ 
		 
		Remark: Notice that the partition-subsets are enumerated with binary\\[-0.08in] 
		representations. These representations carry information about the \emph{path} \\[-0.08in] 
		down the tree which is taken to reach the subset from the start point at $B$.\\[-0.08in] 
		In particular, a `0' indicates \emph{branching} in the leftward direction, whereas `1'\\[-0.08in]
		denotes branching rightwards. For example, the subset $B_{100}$ denotes\\[-0.08in] 
		branching right from $B \rightarrow B_1$ followed by branching left
		from $B_{1 } \rightarrow B_{10}$\\[-0.08in] 
		and finally branching left again from $B_{10} \rightarrow B_{100}$. \\[0.1in] 
		
		\normalsize 3. Specify an infinite set of non-negative numbers: \\[-0.05in] 
		 $\quad \sA = \big \{ \alpha_0, \, \alpha_1, \alpha_{00}, \, \alpha_{01}, \alpha_{10}, \, \alpha_{11}, \, \ldots \big \}$. \\[0.1in] 
		
		\normalsize 4. Attach random probabilities at all edges on the tree. For every binary\\[-0.08in]  
		representation $\epsilon$, draw independently
		 $ \theta_{ \epsilon } \sim \mbox{Beta} (\alpha_{\epsilon0}, \alpha_{\epsilon1}).$ \\
		 
		\normalsize 5. Set the probability of each partition-subset as the product of the\\[-0.08in] 
		probabilities along the path taken. For example, $p(B_{01}) = \theta (1 - \theta_0)$.\\[0.1in] 
		
		\hline
	\end{tabular}
	\\[0.1pt]
\end{table}
\egroup

Following the P\'{o}lya tree prior construction scheme, 
we specify our priors for the unknown distributional components of variable $j$ as
\begin{equation*}
F_{j1}, F_{j0}, F_{j} \sim PT(\Pi_j, \sA_j),
\end{equation*}
for some collections of partition-subsets $\Pi_j$ and non-negative numbers $\sA_j$.

\begin{figure}[h!]
	\centering
	\includegraphics[width=0.6\textwidth]{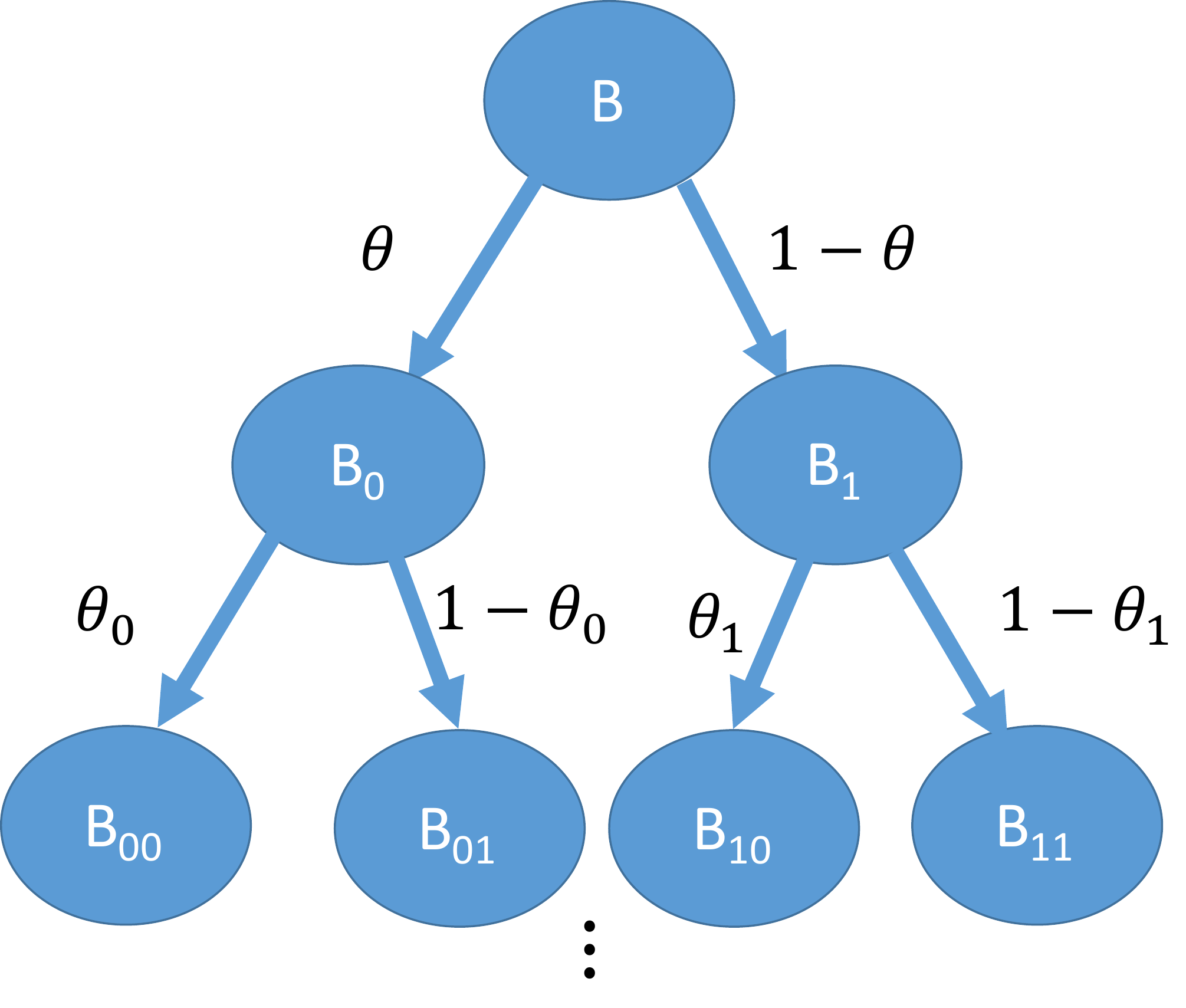}
	\caption{A dyadic tree}
	\label{fig::PTree}
\end{figure}

We consider the setting of hyperparameters $\Pi_j$ and $\sA_j$. Their choices
should depend on the domain and prior information about the unknown distributional components of variable $j$.
Since we are considering only continuous variables here, we ensure that each of the unknown distributional 
components is continuous with probability one \citep{Blackwell1973} by
adopting the canonical choice for $\sA_j$ \citep{Lavine1992} as follows. Let {\tt bin} denote
the set of all binary representations. For any $\epsilon \in \mbox{{\tt bin}}$, 
set
\begin{align}
\label{alphaset}
\alpha_{j, \epsilon 1} = \alpha_{j, \epsilon 0} = \left\{ \begin{array}{ll}
1, & \mbox{ if $l = 0$; and} \\ [2ex]
c_j \times l^2, & \mbox{ if $l \ge 1$},
\end{array} \right.
\end{align}
where $l$ is the length of $\epsilon$ and $c_j$, $1 \le j  \le p$ are smoothing parameters.

We have allowed the smoothing parameters to vary between variables This differs
	from other papers \citep{Hanson2006, Cipolli2016} that deal with multivariate data in which a common smoothing parameter $c$ 
	was used instead of
	a separate smoothing parameter $c_j$ for each variable $j$. Although such a specification
	incurs extra computational cost, we found it to be a necessary price to pay as it takes into account
the varying closeness between the distribution of each variable to its respective P{\'o}lya tree centring distribution
which we shall describe here.

The partitions $\Pi$ may be specified such that the  P\'{o}lya tree prior is centred about a (fixed) ``centring'' distribution $G$, i.e.,
$\bE \{F(A) \} = G(A)$ for all measurable subsets $A \subseteq B$. 
This is implemented through the following scheme - for any binary representation $\epsilon$,
\begin{equation}
\label{partitioning}
B_{\epsilon} = 
\left( 
G^{-1} \left\{ \frac{1}{2^\ell} \sum_{h=1}^{\ell} \epsilon_{h} 2^{\ell-h} \right\}, 
G^{-1} \left\{ \frac{1}{2^\ell} \left( 1 + \sum_{h=1}^{\ell} \epsilon_{h} 2^{\ell-h} \right) \right\} 
\right], 
\end{equation}

\noindent 
where $\ell$ is the length of $\epsilon$, the binary digit $\epsilon_h =0$ if the path involves branching left at layer $h$ of the tree; and $\epsilon_h=1$ otherwise.
We denote a P\'{o}lya tree prior centred about $G$ by $PT(\Pi (G), \sA)$. In our model, we set $\Pi_j = \Pi(G_j)$,
where $G_j$ is the Gaussian distribution parameterised by sample moments as were similarly specified in \cite{ChenHanson}, \cite{Holmes2015},
	and \cite{Cipolli2018}. This is in line with other DA models that outrightly assume a Gaussian distribution for the variables.
However, unlike these Gaussian DA models, the choice of $G_j$ has little influence on the posterior 
inference as it will be overruled by a sufficient amount of data.

\section{Model inference}
\label{Inference}
The objective of our model is to identify the discriminative variables and classify new observations. Suppose we have 
observed $\sY = \{ (y_i, \vx_i), 1 \le i \le n \}$ as realisations
of the model in
Section \ref{Model}. Let $\{(y_{n+r}, \vx_{n+r}), 1 \le r \le m \}$ denote $m$ new observations to be classified.
The required posterior is
\begin{align}
\label{posteriorDensity}
p(\vgamma, \vy^{(new)} \,  | \,  \vx, \vy, \vx^{(new)})
&= \frac{ p(\vx, \vy, \vx^{(new)}, \vy^{(new)}, \vgamma)}{  p(\vx, \vx^{(new)}, \vy) },
\end{align}
where $\vy^{(new)} = (y_{n+1}, \ldots, y_{n+m})$ and $\vx^{(new)} = [\vx_{n+1}, \ldots, \vx_{n+m}]^T$  are the responses and variables of the new observations respectively. 
Clearly, this posterior is
intractable as it requires a sum of $p(\vx, \vy, \vx^{(new)}, \vy^{(new)}, \vgamma)$ over $2^{p+m}$ combinations of $(\vgamma, \vy^{(new)})$.

We will utilise the collapsed variational Bayes (CVB) as described in \cite{tehLDA}
to approximate the required posterior.
This choice of posterior inference stems from the scalability of CVB to high dimensional problems. Since
we have performed posterior inference with variational inference, we shall name our model the 
\emph{variational nonparametric discriminant analysis} (\texttt{VNPDA}).

The CVB approach approximates the actual posterior in (\ref{posteriorDensity}) with a product of densities of the form
\begin{equation*}
q(\vgamma, \vy^{(new)}) = \prod_{r=1}^m q(y_{n+r}) \prod_{j=1}^p q(\gamma_j)
\end{equation*}
that minimises the KL-divergence
\begin{equation*}
\bE_q \left [ \ln \left \{ \frac{q(\vgamma, \vy^{(new)})}{p(\vx, \vy, \vx^{(new)}, \vy^{(new)}, \vgamma)} \right \} \right ],
\end{equation*}
where $\bE_q$ is an expectation with respect to $q(\vgamma, \vy^{(new)})$.
The product components of this optimal $q$-density may be obtained via the iterative equation:
\begin{equation}
\label{prodComp1}
q(\gamma_j) \propto \exp \left [  \bE_{-q(\gamma_j)} \bigg \{ \ln p(\vx, \vy, \vx^{(new)}, \vy^{(new)}, \vgamma) \bigg \} \right ], \quad \mbox{and}
\end{equation}
\begin{equation}
\label{prodComp2}
q(y_{n+r}) \propto \exp \left [ \bE_{-q(y_{n+r})} \bigg \{ \ln p(\vx, \vy, \vx^{(new)}, \vy^{(new)}, \vgamma)  \bigg \} \right ],
\end{equation}
where the expectations are taken with respect to $\prod_{s \neq j} q(\gamma_s) \prod_{r=1}^m q(y_{n+r})$ and $\prod_{j=1}^{p} q(\gamma_j) \prod_{h \neq r} q(y_{n+h})$
respectively. 

Since $\gamma_j$ and $y_{n+r}$ are binary random variables, their respective q-densities
are Bernoulli probability mass functions and hence we need only to compute their
probabilities $\omega_j = q(\gamma_j = 1)$ for $1 \le j \le p$, and $\psi_r = q(y_{n+r} = 1)$ for $1 \le r \le m$.
These values are computed with a coordinate ascent algorithm \citep{Blei2017} as described in Table \ref{Algo1}.
Details on the derivations of the algorithm may be found in \ref{Table1derivations}.
 
The update equations are as follows.
\begin{itemize}
	\item {\bf Update for $\omega_j$}.
	Following equation (\ref{prodComp1}), the resultant update equation for $\omega_j$ is approximately
	\begin{equation}
	\label{approxomega}
	\omega_j
	\approx \expit \left \{ \ln \mbox{BF}_j + \ln (1 + \vone^T \vomega_{-j} ) - \ln (p^u + p - \vone^T \vomega_{-j} - 1)   
  \right \},
	\end{equation}
	where $\vomega_{-j} = \bE_{-q_j} (\vgamma_{-j})$, the column vector $\vgamma_{-j}$ is the result of removing 
	the $j$th entry from $\vgamma$, the function $\expit(z) = (1 + e^{-z})^{-1}$ and
	the term $\ln \mbox{BF}_j$ (refer to equation (\ref{approxlogBF}) for explicit expression) is the log Bayes factor of a two-sample 
	Bayesian nonparametric test between the hypotheses $\{H_{1j}: \gamma_j = 1 \}$ against $\{H_{0j}: \gamma_j = 0 \}$ \citep{Holmes2015}.
	Consequently, $\omega_j$ is an increasing function of a 
	penalised log Bayes factor.
\end{itemize}

To keep the computational cost of the CVB algorithm within an acceptable range, we used the following approximations: (i) 
Taylor's expansion on nonlinear functions of $\vone^T \vgamma_{-j}$, e.g. $\ln(1 + \vone^T\vgamma_{-j})$;
(ii) Stirling's approximation \citep{Abramowitz2002} on beta functions involving $y_{n+r}$ (under the assumption that $1/n$ is small).
The reader may refer to \ref{Table1derivations} for details on how we have applied the approximations\wyc{.}


Observe that the penalty term 
$\ln (1 + \vone^T \vomega_{-j} ) - \ln (p^u + p - \vone^T \vomega_{-j} - 1)$
is decreasing in $u$. Therefore, the setting of $u$ should be considered carefully
as it controls the trade-off between errors of type I (selecting truly non-discriminative variables as discriminative) 
and type II (missing out on truly discriminative variables).

\begin{itemize}
	\item {\bf  Update for $\psi_r$}.
	The update equation for $\psi_r$ is approximately
	%
	\begin{align}
	\label{approxpsi}
	\psi_r
	&\approx \expit \left [  \ln \bigg ( \frac{a_y + n_1}{b_y + n_0} \bigg )
	+ \vomega^T  \ln \big \{ \vpi_r^{(1)} \big \} 
	- \vomega^T \ln \big \{ \vpi_r^{(0)} \big \} \right ],
	\end{align}
	
	\noindent where the number of observations from response group $k$ is $n_k$,
	the column vector $\vpi_r^{(k)}$ of size $p$ is such that
	the j-th element is
	\begin{align*}
	\pi_{rj}^{(k)} &\approx \exp\Bigg [ \sum_{ \ell \le N_j } \bigg \{ \ln \Big \{ \alpha_{j, \epsilon_{rj}(\ell+1)} + n_{j, \epsilon_{rj}(\ell+1)}^{(k)} \Big \} \\
	&\hspace{2.5cm} - \ln \Big \{ 2\alpha_{j, \epsilon_{rj}(\ell+1)} + n_{j, \epsilon_{rj}(\ell) }^{(k)} \Big \} \bigg \} \Bigg ],
	\end{align*}
	the binary representation 
	$\epsilon_{rj}(\ell)$ denotes the first $\ell$ branching directions of the path of $x_{n+r,j}$ through the P\'{o}lya tree $PT(\Pi (G_j), \sA_j)$, 
	the number of observations from response group $k$
	that fall in the partition-subset $B_{j, \epsilon }$ is  $n_{j, \epsilon }^{(k)}$, and $N_j$ is
	a constant. 
	
\end{itemize}

\bgroup
\def\arraystretch{1.4}
\begin{table}[ht!]
	\refstepcounter{table}\label{Algo1}
	\centering
	\begin{tabular}{l}
		\normalsize
		\bf{Table 2}~ \normalfont{Iterative scheme for obtaining the parameters of the optimal } \\ densities $q(\vgamma, \vy^{(new)})$.  \\
		\hline
		\normalsize Require: For each $j$, initialise $\omega_j^{(0)}$ with a number in $(0,1)$.\\
		\normalsize \textbf{while} $||\vomega^{(t)} - \vomega^{(t-1)}||^2$ is greater than tolerance $\tau$,  \textbf{do} \\
		\normalsize \qquad At iteration $t$, compute for $j =1, \ldots, p$, \\
		\normalsize \qquad 1: $ \eta_j^{(t)} \leftarrow  \ln \mbox{BF}_j + \ln \bigg \{1 + \vone^T \vomega_{1:j-1}^{(t)} + \vone^T \vomega_{j+1:p}^{(t-1)} \bigg \}$\\ [2ex]
		\hspace{3.3cm} $- \ln \bigg \{p^u + p- \vone^T \vomega_{1:j-1}^{(t)} - \vone^T\vomega_{j+1:p}^{(t-1)} - 1 \bigg \}$ \\
		\normalsize \qquad 2: $\omega_j^{(t)} \leftarrow  \text{expit} (\eta_j^{(t)})$ \\
		\normalsize Upon convergence of $\vomega$, compute for $r = 1, \ldots, m$, \\
		\normalsize 3: $\psi_r \leftarrow \expit \left [ \ln (a_y + n_1) - \ln (b_y + n_0)
		+ \vomega^T  \ln \big (\vpi_{r}^{(1)} \big ) 
		- \vomega^T \ln \big (\vpi_{r}^{(0)} \big ) \right ] $\\[0.15in]
		\hline
	\end{tabular}
	\\[0.1pt]
\end{table}
\egroup

Both Taylor's and Stirling's approximations are also used to obtain (\ref{approxpsi}). 
This allows us to update $\vomega$ in isolation before using the converged value to calculate each $\psi_r$ 
individually, thus reducing the computational cost.

The final classification rule may be heuristically interpreted as a function of
pseudo-proportion ratios. Observe that the term 
\begin{equation*}
\vomega^T  \ln \big \{ \vpi_r^{(1)} \big \} 
- \vomega^T \ln \big \{ \vpi_r^{(0)} \big \} = \sum_{j=1}^p \omega_j \left \{ \ln (\pi_{rj}^{(1)} / \pi_{rj}^{(0)})  \right \},
\end{equation*}
is a weighted sum of log proportion ratios. Briefly, the proportion
$\pi_{rj}^{(1)}$ is large if a large proportion of observations from group 1 have
similar branching directions as $x_{n+r,j}$. In other words, the classification rule classifies a new observation as
group 1 if its path is more similar to paths taken by group 1 observations than those from group 0.

	Our approach to computing the  marginal log-likelihood of the data 
	is very similar to that described by \cite{Holmes2015}.
	The computational efficiency of the algorithm is of order $O(np)$ based on the settings justified in
	\ref{Table1derivations}. This comes from traversing through $p$ P{\'o}lya trees each truncated at layer
	$\lfloor \ln_2 (n) \rfloor$.

\section{The smoothing parameter $c_j$}
\label{ChooseC}

The choice of 
the $c_j$'s is crucial as it affects both the variable selection and classification rules. Unfortunately, 
most existing options in the literature such
as assigning a hyper-prior and empirical estimation have been proposed 
in the context of low dimensional problems
only and are not easily scalable to high dimensionality settings.
\cite{Hanson2002}, \cite{Zhao2011}, and \cite{Cipolli2016} dealt with a limited number of 
smoothing parameters in their respective models by assigning them a hyper-prior. 
Although these methods yielded good numerical results in their papers, 
putting hyper-priors on the $c_j$'s will compound the computational difficulties we already have to tackle.
More specifically, a hyper-prior on each $c_j$ 
will lead to an extra $2p$ update steps in our coordinate
ascent algorithm (see Table 2). These extra steps may be bypassed by collapsing our
model over the $c_j$'s but it will lead to difficulties with the computation of the $q$-densities
for $\gamma_j$ and $y_{n+r}$.
\cite{Holmes2015} found that any value of a single smoothing parameter $c$ 
between 1 to 10 works well in practice. However, a
generalisation to multiple smoothing parameters has not been discussed.
Among the empirical estimation methods \citep{Berger2001, ChenHanson, Holmes2015, Cipolli2018}, the only algorithm that
has been applied to a model
with multiple $c$'s has been proposed in \cite{ChenHanson}.
However, this approach involves a $p$-dimensional grid search
in the high-dimensionality context which is infeasible in our context.  

Since our context is the analysis of high dimensional data, we suggest a novel heuristic approach of choosing
$c_j$ based on an \emph{a priori} analysis
that can be executed efficiently. More specifically, we ``infer" a likely value of $c_j$ from a list of 
candidate values that has generated our observed data
under each of the two possible hypotheses $\{H_{0j}: \gamma_j = 0\}$ and $\{H_{1j}:\gamma_j = 1 \}$.
Under $H_{0j}$, the value of $c_j$ that generated our data is likely to be large if the empirical 
distribution of $F_j$ is close to $G_j$, whereas under $H_{1j}$, 
the value of $c_j$ that generated our data is likely to be large if the Euclidean distance between 
the empirical distributions of $F_{j1}$ and $F_{j0}$ is small. 
To make the implementation more computationally feasible, the variables are grouped into clusters 
such that variables within each cluster
are assumed to have equal values of $c_j$. The best setting for the $c_j$'s are selected to
minimise resubstitution classification error. This choice of objective function 
aligns with the classification objective of our proposed model and differs from the log marginal likelihood objective
used in \cite{ChenHanson}. Detailed steps
of this proposed \emph{a priori} analysis can be found in Table \ref{Algo2}.


\bgroup
\def\arraystretch{1.2}
\begin{table}[ht!]
	\refstepcounter{table}\label{Algo2}
	\centering
	\begin{tabular}{l}
		\normalsize
		\bf{Table 3}~ \normalfont{Method for choosing } $c_1, \ldots, c_p$ \\
		\hline
		\normalsize 1. Under $H_{0j}$, we assess the goodness of fit of $F_j$ to $G_j$ for each $j$ by the\\[-0.05in] 
		p-value of an appropriate hypothesis test, eg. the Shapiro-wilk's test.\\[-0.05in]  
		Here, a large p-value favours a large $c_j$.\\
		\normalsize 2. Under $H_{1j}$, we assess the distance between $F_{j1}$ and $F_{j0}$ by computing\\[-0.05in]  
		the p-value of a Kolmogorov-Smirnov test. Here, a large p-value favours\\[-0.05in] 
		a large $c_j$.\\
		\normalsize 3. Denote the p-values computed in steps 1 and 2 by $\{v_{j0}\}_{j=1}^p$ and $\{v_{j1}\}_{j=1}^p$\\[-0.05in] 
		respectively. Let $\tilde{v}_j$ be the p-value of whichever $H_{0j}$ or $H_{1j}$ is true, i.e.\\[-0.05in]  
		$\tilde{v}_j = v_{j0}$ if $H_{0j}$ is true; $\tilde{v}_j = v_{j1}$ otherwise. Calculate the prior expected\\[-0.05in] 
		value of $\tilde{v}_j$:\\
		\hspace{3cm} $E_j = \bE(\tilde{v}_j) = (v_{j1} + p^u \times v_{j0})/(1 + p^u)$.\\[0.1in]

		\normalsize 4. Rank the expected values in ascending order: $E_{(1)} , \ldots, E_{(p)}$. \\[-0.05in]
		\normalsize 5. Assign 
		
		\hspace{1cm} \begin{minipage}{4cm}\begin{equation*} 
			c_j = \left\{ \begin{array}{ll}
			a_1, & \mbox{ if $E_j < E_{(\lfloor p/4 \rfloor )}$;} \\ [1ex]
			a_2, & \mbox{ if $E_{(\lfloor p/4 \rfloor )} \le E_j < E_{(\lfloor p/2 \rfloor )}$;} \\ [1ex]
			a_3, & \mbox{ if $E_{(\lfloor p/2 \rfloor )} \le E_j < E_{(\lfloor 3p/4 \rfloor )}$;} \\ [1ex]
			a_4, & \mbox{ if $E_j  \ge E_{(\lfloor 3p/4 \rfloor )}$,}
			\end{array} \right.
			\end{equation*}\end{minipage}\\[0.5in] 
		where $a_1 \le a_2 \le a_3 \le a_4$ are constants that may be chosen from the range\\[-0.05in] 
		$(0, 100]$ to minimise resubstitution classification error in the training data. \\
		
		\hline
	\end{tabular}
	\\[0.1pt]
\end{table}
\egroup

\section{Numerical results}
\label{Numerical}
In this section, we examine the performance of our proposed method
in 6 simulation settings and 2 publicly-available gene expression datasets.
We fitted our proposed \texttt{VNPDA} model with the publicly available \texttt{R} package that can be found at the website
\texttt{https://github.com/weichangyu10/VaDA}
and compared its performance with the classifiers - variational linear discriminant analysis \citep[\texttt{VLDA},][]{Yu}, 
variational quadratic discriminant analysis \citep[\texttt{VQDA},][]{Yu}, 
penalised-LDA \citep[\texttt{penLDA},][]{WittenTibs}, nearest shrunken centroid \citep[\texttt{NSC},][]{TibshiraniNSC} (\texttt{NSC}) 
and na{\"i}ve Bayes kernel discriminant analysis \citep[\texttt{naiveBayesKernel},][]{Strbenac2015}. 
Both \texttt{VLDA} and \texttt{VQDA} are Bayesian analogues of 
na{\"i}ve Bayes discriminant analysis \citep{McLachlan1992} that have exhibited competitive 
classification errors. 
The \texttt{penLDA} classifier is a penalised version of 
Fisher's discriminant analysis that performs well when the true signal (difference in true group-conditional
means divided by standard deviation) is sparse, 
whereas the \texttt{NSC} classifier has been chosen 
as a competing classifier due to its popularity in bioinformatics literature.
Lastly, the \texttt{naiveBayesKernel} is a two-stage 
nonparametric classifier that performs variable selection
with the Kolmogorov-Smirnov test before fitting the selected variables into a {na\"ive} Bayes discriminant analysis
model that estimates the group-conditional distributions with kernel density estimates. 
Further details of the comparison methods have
been provided in Section 1 of the Supplementary Material.

\subsection{Simulation Study}
\begin{figure}[h!]
	\centering
	\includegraphics[width=\textwidth,keepaspectratio]{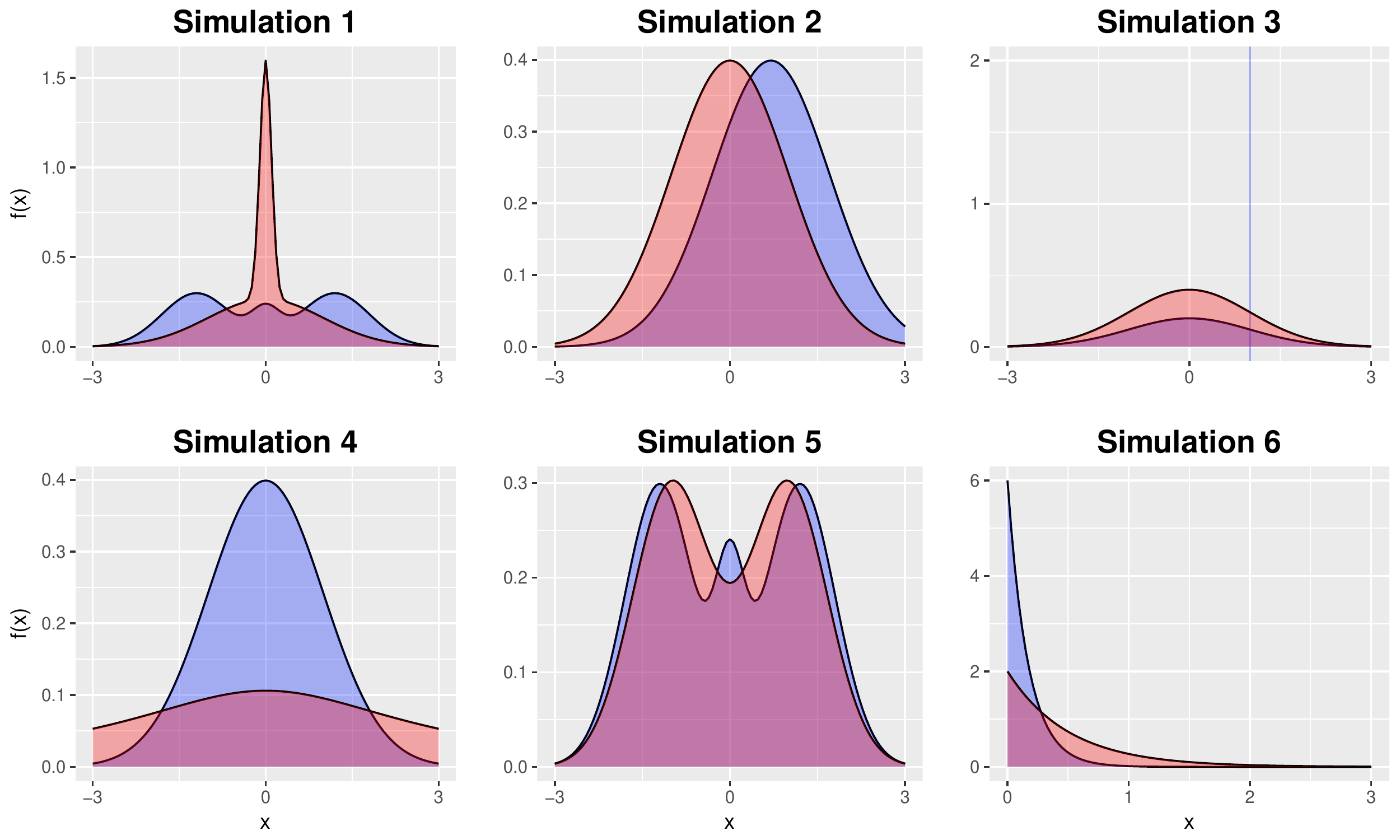}
	\caption{Distribution of discriminative variables.}
	\label{fig::SimSetting}
\end{figure}

The models are trained with 
$n= 100$ observations and are used to classify
$m= 1000$ new observations in each simulation setting for $50$ repetitions. 
At each repetition, the simulated dataset consists of $50$ 
truly discriminative variables that follow
various non-Gaussian distributions (simulation 2 as an exception), and 450 non-discriminative variables. Details
of their distributions are provided below and in Figure \ref{fig::SimSetting}. 

\medskip
\noindent \underline{Simulation 1} \\
We compare the models' performances in discriminating a trimodal distribution from a kurtotic unimodal distribution. The distributions have equal means. These two distributions are mentioned in \cite{MarronWand}. \\ 
Group 1:  $(9/20) \sN ( - 6/5, (3/5)^2 ) + (9/20) \sN ( 6/5, (3/5)^2 ) + (1/10) \sN ( 0, (0.25)^2 )$.
Group 0: $(2/3) \sN ( 0, 1 ) + (1/3) \sN ( 0, 0.1^2 )$.  

\medskip 
\noindent \underline{Simulation 2} \\
Here, we assess the loss incurred by nonparametric classification when the Gaussian assumption holds. \\ Group 1: $\sN (0.7, 1)$. \\
Group 0: $\sN (0, 1)$. 

\medskip 
\noindent \underline{Simulation 3} \\
We examine the models' ability to discriminate distributions that differ by a density spike
at $x=0.5$. \\ 
Group 1: $0.5 \, \sN (0, 1) + 0.5 \, \sN (0.5, 0.001^2) $. \\
Group 0: $\sN (0, 1)$.  

\medskip 
\noindent \underline{Simulation 4} \\
We assess the models' performances when the group-conditional distributions differ largely by tail thickness. \\ Group 1: $\sN (0, 1)$. \\
Group 0: Cauchy distribution with location $0$ and scale $3$.   

\medskip
\noindent \underline{Simulation 5} \\
Here, we compare the models' performances in a challenging classification scenario when the group-conditional distributions differ by an additional minor mode between two major modes. These two distributions are mentioned in \cite{MarronWand}. \\ Group 1: $(9/20) \sN ( - 6/5, (3/5)^2 ) + (9/20) \sN ( 6/5, (3/5)^2 ) + (1/10) \sN ( 0, (0.25)^2 )$. \\
Group 0: $0.5 \sN (-1, (2/3)^2) + 0.5 \sN (1, (2/3)^2) $.   

\medskip
\noindent \underline{Simulation 6} \\
We assess the models' performances when discriminating two Exponential distributions of different rates. \\ Group 1: Exp$(6)$. \\ 
Group 0: Exp$(2)$. 

\medskip
The non-discriminative variables are partitioned into 9 groups of 50 variables.
Within each group, variables are independent, identically distributed. The distributions
of each group are:
  $t_1$, $\mbox{Cauchy}(0,2)$, $\mbox{Gamma}(2,2)$, $\mbox{Exp}(1)$, $\sN(0, 5^2)$, 
$\sN(0,1)$, $0.1 \sN (0,1) + 0.9 \sN(0,0.1^2)$ (zero-inflated model), \\
$\sum_{\ell = 0}^{7} (1/8)\sN (3 \{ (2/3)^{\ell} - 1 \},  (2/3)^{2\ell} )$ (multiple modes), and \\
$0.5 \sN (-1.5,0.5^2) + 0.5 \sN(1.5,0.5^2)$ (bi-normal).

\begin{figure}[h!]
	\centering
	\includegraphics[width=\textwidth,keepaspectratio]{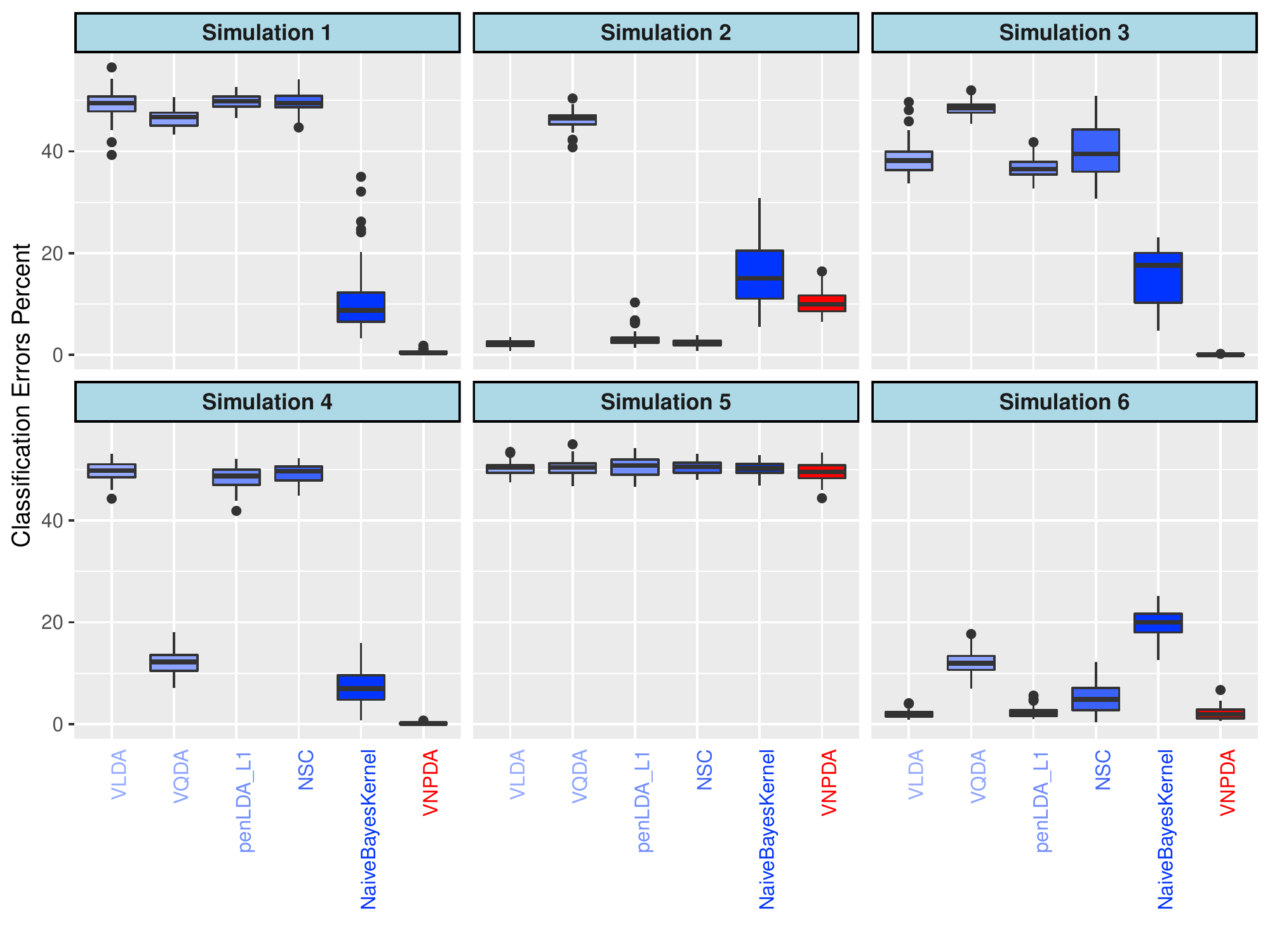}
	\caption{\normalsize Classification errors of simulation study for $n=100$ training and $m=1000$ new observations.}
	\label{fig::SimClassErrors}
\end{figure}

\begin{figure}[h!]
	\centering
	\includegraphics[width=1.2\textwidth,keepaspectratio]{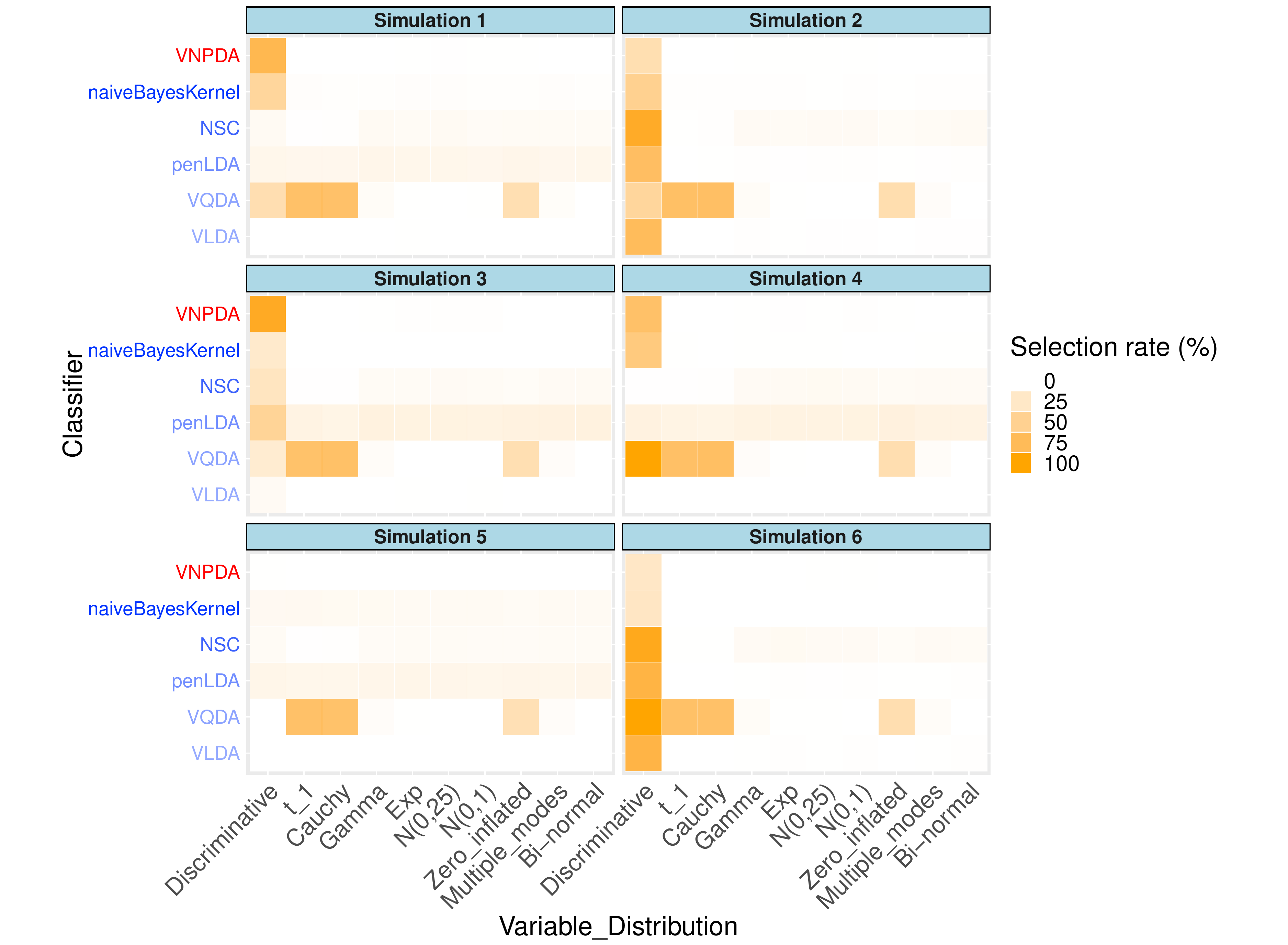}
	\caption{Variable selection rates by simulation setting for $n=100$ training observations.}
	\label{fig::SimAllSelectRate}
\end{figure}

\smallskip
Results of the simulation study are summarised in 
Figure \ref{fig::SimClassErrors}, Figure \ref{fig::SimAllSelectRate}, and Table \ref{Accuracy}. The median computation time
of \texttt{VNPDA} for one repetition is approximately 55s (an acceptable time) on a 1.6GHz Intel Core i5 processor.

In simulations 1 and 4, three models \texttt{VNPDA}, \texttt{naiveBayesKernel} and \texttt{VQDA} have 
high selection rates among 
the discriminative variables. In simulation 1, the group-conditional distributions have well-separated major
modes ($0$ vs. $\pm 6/5$), while the two distributions have differing tail thickness in simulation 4. However,
\texttt{VQDA} did not perform as well in variable selection accuracy as it is unable to distinguish noise generated by 
non-discriminative variables 
with thick-tailed distributions such as $t_1$ and Cauchy$(0,2)$. This directly leads to \texttt{VQDA}'s poor classification performance. 
\texttt{VNPDA} performed excellently 
in simulation 3 and this is evidence
of its superiority in detecting density spikes. However, it did not perform better than the other models in 
simulations 2 and 6. In simulation 2, we expected the nonparametric classifiers \texttt{naiveBayesKernel}
and \texttt{VNPDA} to exhibit poorer performance than the other models as the Gaussian assumption holds. In simulation 6, \texttt{VLDA} and \texttt{penLDA} exhibited
a slight edge over \texttt{VNPDA} as the nonparametric option did not perform as well in identifying the discriminative variables.
This is due to the lack of separation between the modes of the group-conditional distributions.

We repeated the simulation study for various training sample sizes ($n=30$ and $n=500$) and the results are displayed in
	Section 2 of the Supplementary Material. Classification and variable selection performances generally improve across all models for Simulations 2, 3 and 6 as $n$ increases and remain
	similar for other simulation settings. The classification errors for \texttt{VNPDA} appear
	to have stabilised at $n=100$ for all simulation
	settings except for Simulation 2. When the Gaussian assumption is indeed true,
	nonparametric DA models require a much larger training sample size to achieve
	similar classification error with Gaussian DA models. 
	No significant changes in the performance ranking have been observed as sample size changes. These findings suggest that the methods are 
	sample-efficient in Simulations 2, 3 and 6 and that the 
	results hold for a wide range of sample sizes.

Overall, the conditions which appear to be favourable to \texttt{VNPDA} are: (i) differing tail thickness; 
(ii) well-separated major modes.

\begin{table}[h!]
	\centering
	\begin{tabular}{ ccccccc  }
		\hline
		Sim. Set.& \texttt{VLDA} &\texttt{VQDA} &\texttt{penLDA} &\texttt{NSC} &\texttt{na{\"i}veBayesKernel} &\texttt{VNPDA}\\
		\hline
		Sim. 1 &89.91          &75.72  &83.07 &87.13 &93.10 &{\bf 97.62} \\
		Sim. 2 &{\bf 97.02}    &76.22  &96.58 &95.87 &94.38 &93.36\\
		Sim. 3 &90.40          &74.37  &82.72 &89.48 &92.25 &{\bf 99.09}\\
		Sim. 4 &89.93          &81.84  &79.90 &86.36 &95.55 &{\bf 96.60}\\
		Sim. 5 &89.97          &72.52  &82.69 &87.32 &85.68 &{\bf 90.00}\\
		Sim. 6 &{\bf 97.85}    &81.98  &97.88 &96.12 &92.66 &92.48\\
		\hline
	\end{tabular}
	\caption{Accuracy ($(TP+TN)/(P+N) \times 100$ \%) of variable selection for each simulation setting ($n=100$), where
	$TP$ equals the no. of true positives, $TN$ equals the no. of true negatives, $P$ equals the no. of positives and $N$ equals the no. of negatives.}
	\label{Accuracy}
\end{table}

\newpage
\subsection{Application to gene expression datasets}

\noindent \underline{Melanoma dataset} \\
The melanoma dataset has been analysed by \cite{Mann}. 
The data underwent preprocessing to remove underexpressed genes, i.e. median $\le 7$. 
Patients whose survival time are lesser than 1 year and died due to the disease are labelled as poor prognosis; 
patients who are still alive and free from Melanoma after 4 years are labelled as good prognosis.
Finally, we standardise the dataset to obtain $z_{ij} = (x_{ij} - \overline{x}_j)/s_{j}$, where $x_{ij}$ is the
gene $j$ reading for observation $i$. The processed dataset that has been analysed consists of $n=47$ 
observations by $p=12881$ DNA microarray readings. 

\medskip 
\noindent 
\underline{Sarcoma dataset} \\
The sarcoma dataset is uploaded by \cite{TCGAbiolinks} and is made publicly available on \url{bioconductor}. 
The data underwent preprocessing to retain only variables with variance $> 0.1$.
Patients whose survival time are lesser than the 20th percentile are labelled as poor prognosis; 
patients whose survival time are greater than 80th percentile are labelled as good prognosis.
The dataset is standardised in a similar manner to the melanoma dataset. 
The processed dataset that has been analysed consists of $n=74$ observations by $p=20449$ RNA readings.\\

\begin{figure}[h!]
	\centering
	\includegraphics[width=0.8\textwidth,keepaspectratio]{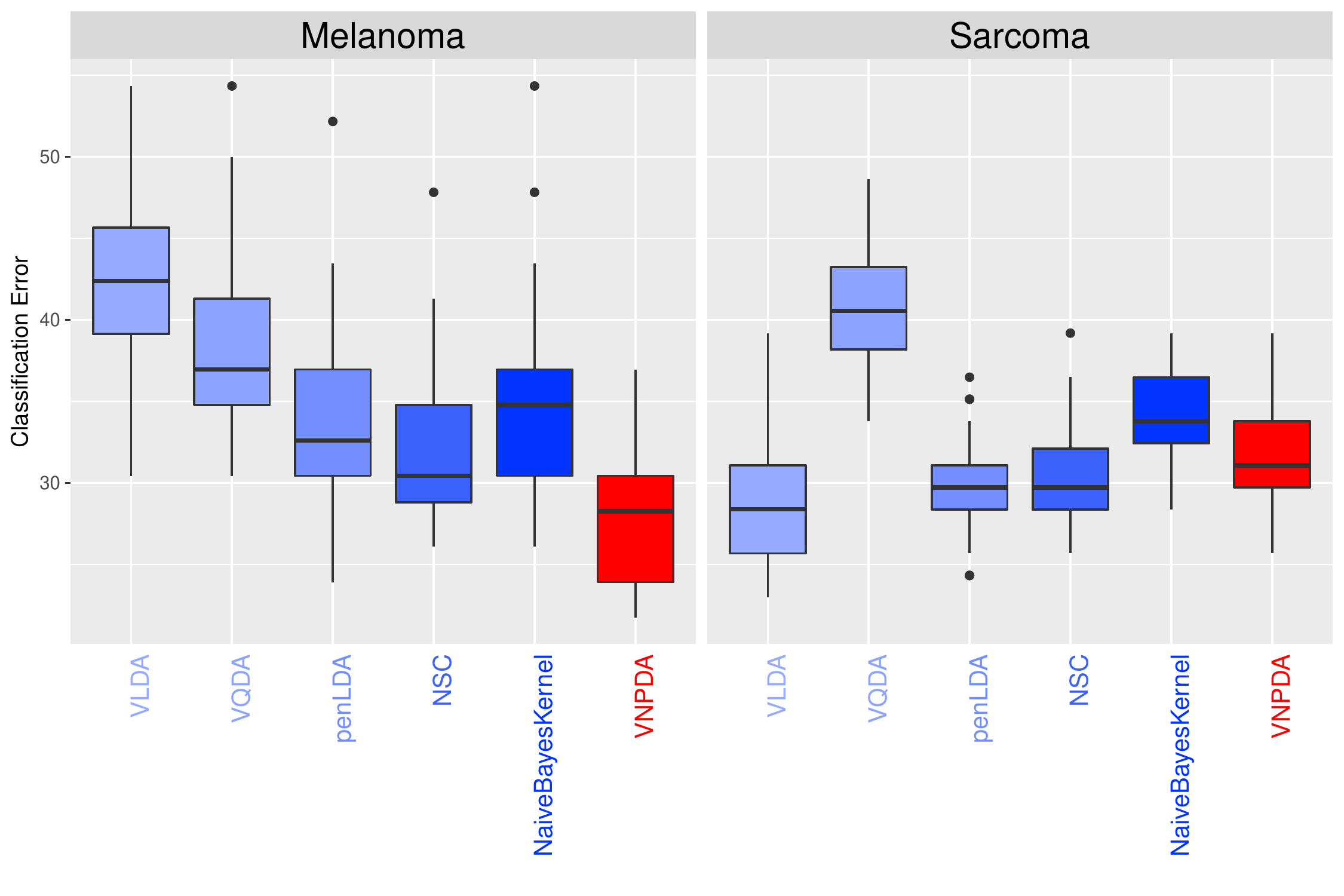}
	\caption{5-fold cross validation classification errors of gene expression datasets.}
	\label{fig::RealDataError}
\end{figure}

The classification errors are summarised in 
Figure \ref{fig::RealDataError}. The median computation times for one CV iteration
are approximately 80s and 200s for the melanoma and sarcoma dataset respectively on a 1.6GHz Intel Core i5 processor. This is slower than other methods but is believed to be
within an acceptable range to most users. In terms of
classification errors, \texttt{VNPDA} outperformed the Gaussian DA models, including \texttt{VLDA}, 
in the melanoma dataset but there is no significant difference in performance in the sarcoma dataset. This warrants a more
detailed investigation into the reasons that led to this disparity. For simplicity, we shall focus on the performances
of \texttt{VLDA} and \texttt{VNPDA} as these classifiers exhibited the greatest contrast in performance between the two gene expression datasets.
A visualisation of genes selected by each classifiers has been provided in Section 3 of the Supplementary Material.

\begin{figure}[h!]
	\centering
	\includegraphics[width=0.95\textwidth,keepaspectratio]{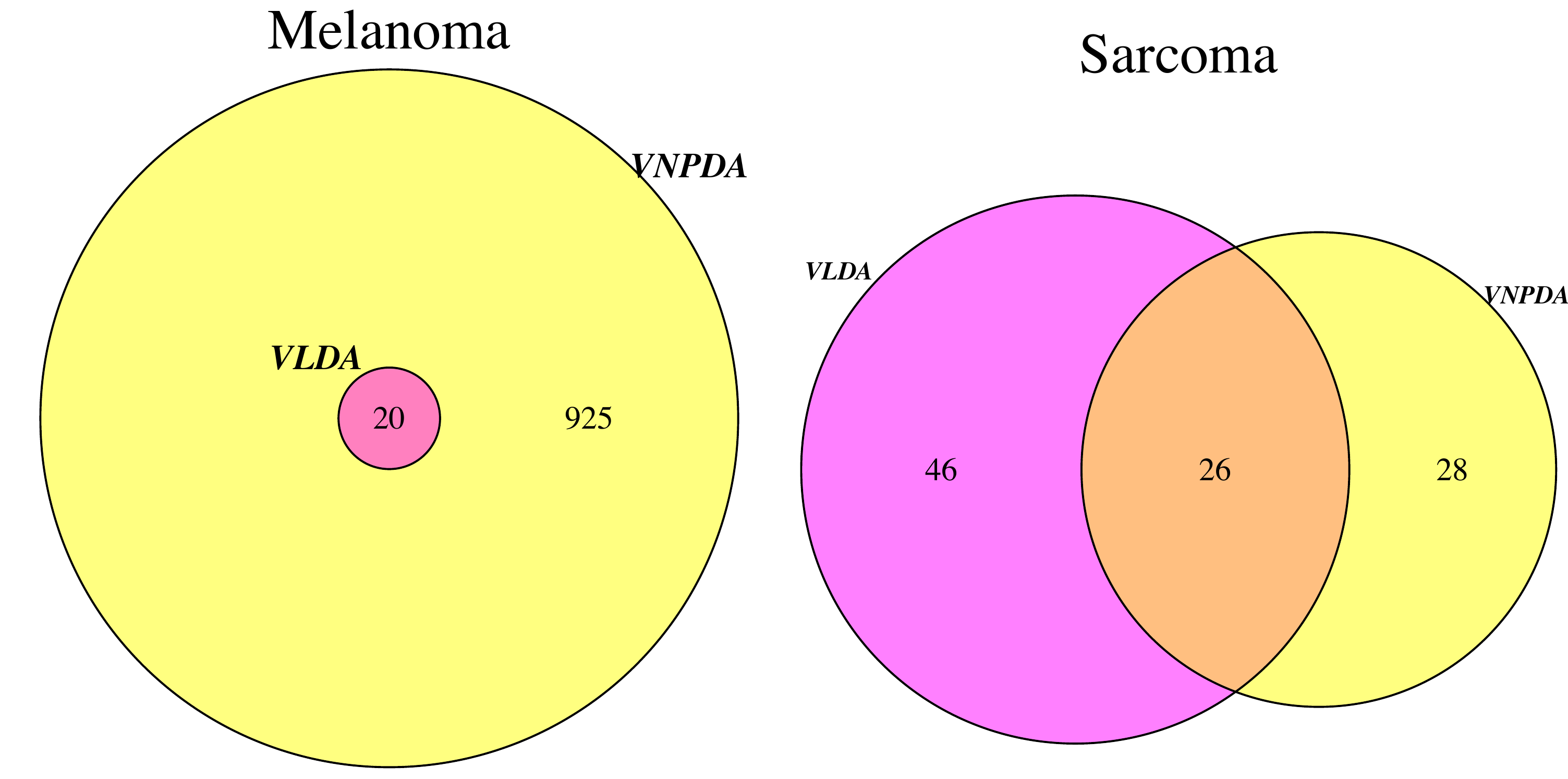}
	\caption{Number of frequently selected genes (selection rate $>$ 20\%).}
	\label{fig::VennSelection}
\end{figure}

Based on the venn diagrams in Figure \ref{fig::VennSelection}, it is clear that the disparity is due to the 
number of frequently selected genes. In particular, we found that
\texttt{VNPDA} has more frequently selected genes than \texttt{VLDA} in the melanoma dataset, whereas \texttt{VLDA} has more frequently
selected genes than \texttt{VNPDA} in the sarcoma dataset.

\begin{figure}[h!]
	\centering
	\includegraphics[width=0.95\textwidth,keepaspectratio]{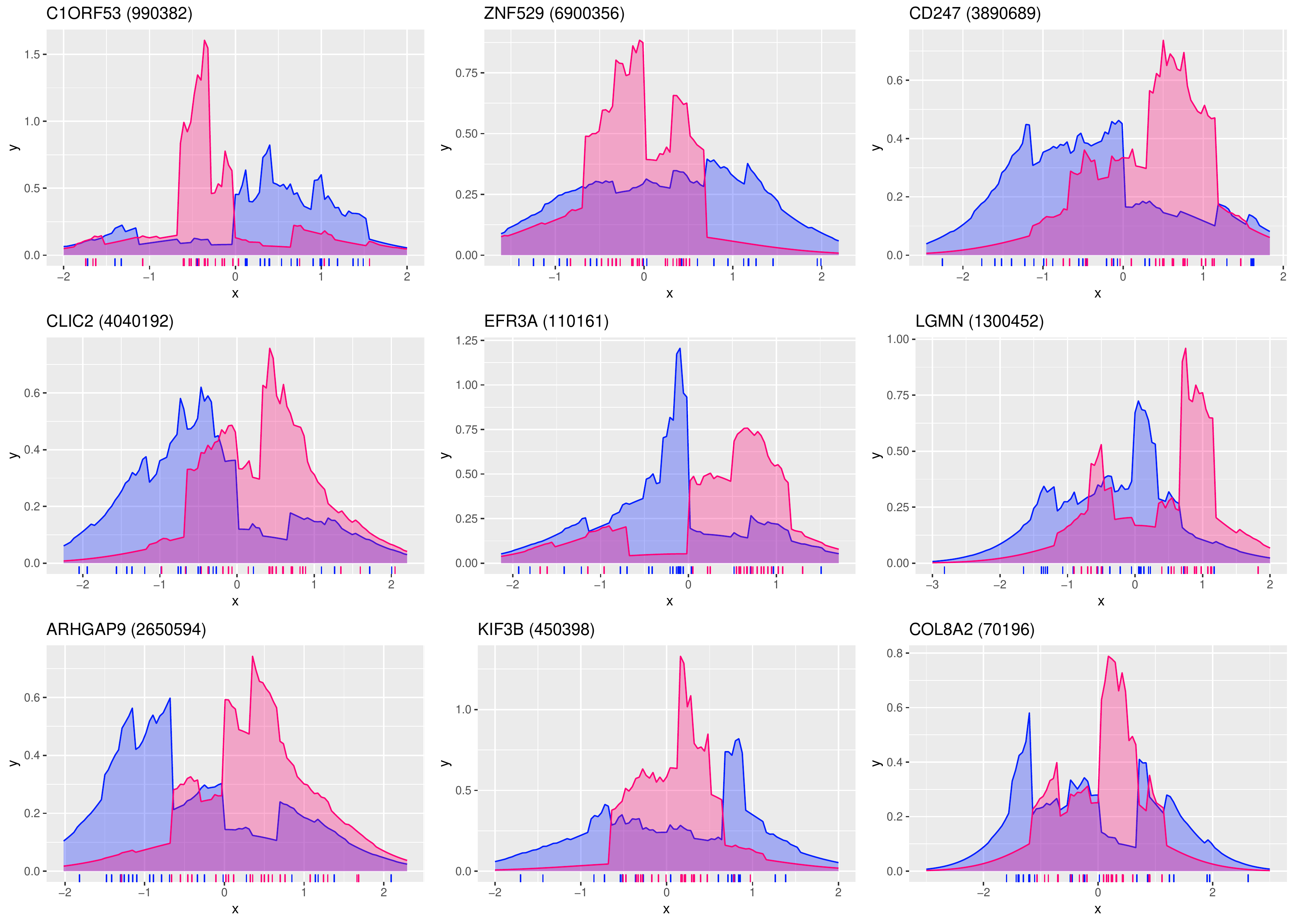}
	\caption{P\'{o}lya tree predictive density ($c$=1) of 9 frequently selected genes by \texttt{VNPDA} but not \texttt{VLDA} for melanoma dataset.}
	\label{fig::PTMelanoma}
\end{figure}

Next, we shall identify the reason that led to this difference in the number of frequently selected genes.
In the melanoma dataset, there are 925 genes selected by
\texttt{VNPDA} but not \texttt{VLDA}. Among these genes, a substantial proportion (18.8\%) are not selected because
their group-conditional distributions exhibited a strong departure from normality assumptions (Shapiro-Wilk's $p$-value $<$ 0.05).
Figure \ref{fig::PTMelanoma} presents the P\'{o}lya predictive density plot \citep[see][for definition]{Hanson2002} of $9$ of these genes which shows that
their group-conditional distributions are favourable for \texttt{VNPDA} to perform well, i.e., they either 
differ in tail thickness or have well-separated major modes. For example, the group-conditional distribution of
the poor prognosis group (red) for the C1ORF53 gene looks like a kurtotic unimodal distribution.

\begin{figure}[h!]
	\centering
	\includegraphics[width=0.95\textwidth,keepaspectratio]{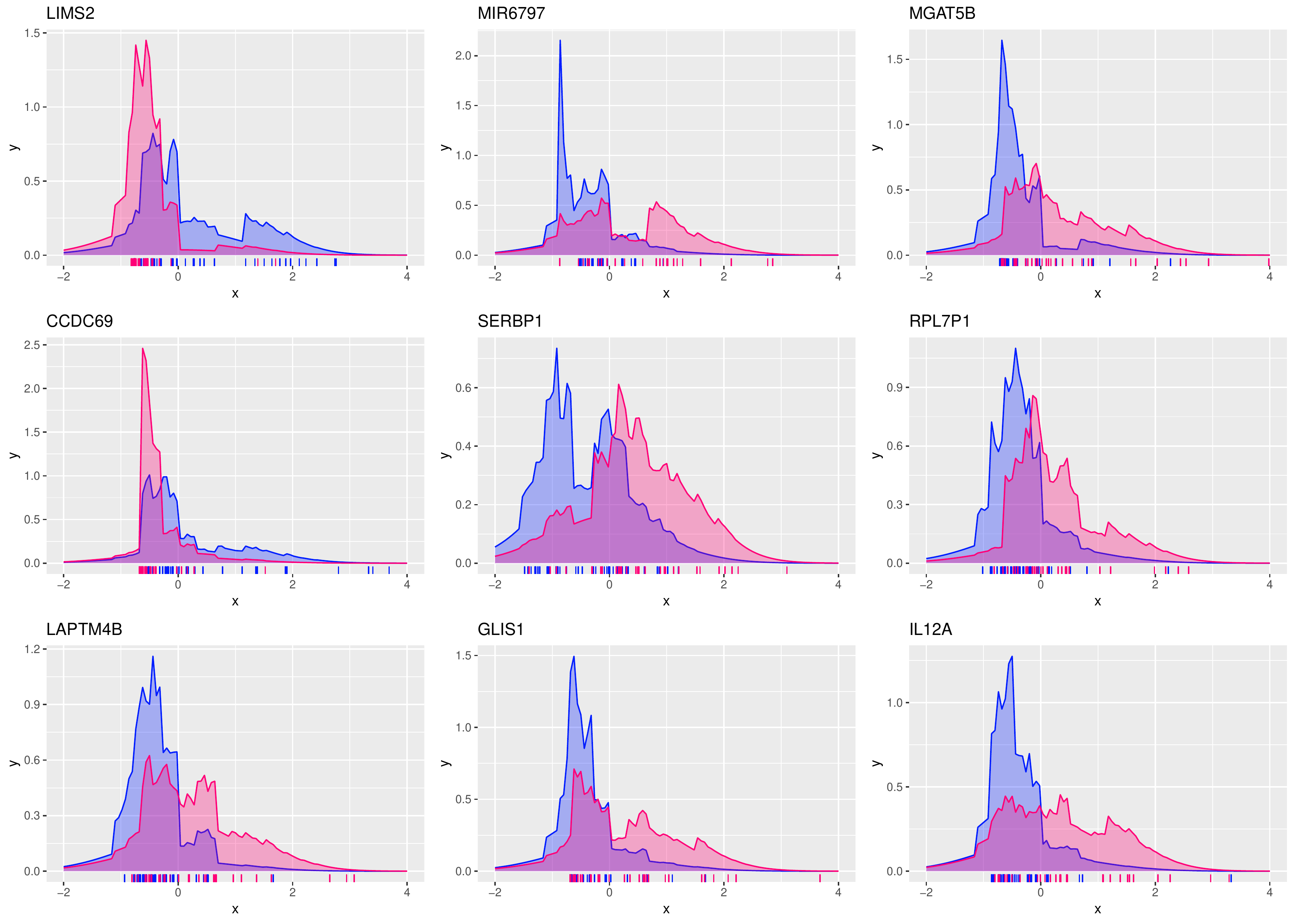}
	\caption{P\'{o}lya tree predictive density ($c$=1) of 9 frequently selected genes by \texttt{VLDA} but not \texttt{VNPDA} for sarcoma dataset.}
	\label{fig::PTSarcoma}
\end{figure}

In contrast, more genes are selected
by \texttt{VLDA} than \texttt{VNPDA} in the sarcoma dataset. By plotting the P\'{o}lya predictive density of genes that are selected by
\texttt{VLDA} but not \texttt{VNPDA}, we observed that
many of these genes have unimodal, right skewed group-conditional distributions (see Figure \ref{fig::PTSarcoma}). 
We notice that
the positions of their group-conditional major modes nearly coincide with one another. Such conditions lead to a smaller Bayes factor for the
\texttt{VNPDA} variable selection rule
and hence weaker evidence for the variable to be discriminative. On the other hand, the \texttt{VLDA} variable selection rule performed
better as they depend only on the separation between the group-conditional means, and these are indeed well-separated among most of the genes.


\newpage
\section{Discussion}
\label{Conclusion}

In this paper, we presented a novel Bayesian discriminant analysis model
that performs both variable selection and classification without 
making assumptions about the parametric form of the unknown distributions.
To deal with our uncertainty about these distributions, we
assigned them with P{\'o}lya tree priors.
Since the results using the P{\'o}lya tree prior are sensitive to the choice 
of the smoothing parameter, we suggested a data-driven approach based on an \emph{a priori} inference
that helps with the choice of this parameter.
By adopting a CVB approximation for posterior inference, we arrive at a classification rule
that carries a heuristic interpretation.

As Bayesian nonparametric methods are unpopular for analysing high-dimensional data due
to the computational cost, we applied computational short-cuts
to the CVB update equations. The approximations
effectively isolate the updates of the variable selection probabilities from the classification probabilities.
Thus, we compute the variable selection probabilities in an iterative loop before using the converged value
for calculating the classification probabilities.
This, in combination with an implementation in \texttt{C++}, led to a computation cost that we believe is within an acceptable range (80 to 200s) for most users
in both the simulated and publicly available datasets
we have examined.

The numerical results indicate that our proposed model performs reasonably well in most cases
and is superior when the group-conditional distribution have either well-separated major modes or differing tail thickness.
The findings are validated when we examine the group-conditional distributions of the variables in two publicly-available datasets.

Our proposed model structure also allows for several possible extensions.
A possible extension that is beyond the scope of the motivating problem
is to accommodate categorical variables in three scenarios. In the first scenario whereby all variables are binary,
we may model each variable with a pair of Bernoulli-beta group-conditional distributions
under the alternative hypothesis and a corresponding
common Bernoulli-beta distribution under the null.
In the second scenario whereby we have a mix of continuous and binary variables, the 
continuous variables may be modelled in accordance to equation (\ref{Model}), while
binary variables are modelled with the Bernoulli-beta distributions. In the third scenario whereby we have a mix of
continuous and nominal variables, each nominal variable may be modelled with a multinomial-Dirichlet distribution.
If the variables are ordinal instead of nominal, we may utilise the structure  of the categorical ordering
by penalising the differences between adjacent categorical probabilities \citep[see for example][]{Gertheiss2009, WittenTibs}.
To reduce the number of parameters to be estimated, the models in all three scenario may be marginalised over the probability parameters.
When there is a need to account for correlation between variables, suitable strategies include
re-representing each multi-category variable with a set of binary variables \citep{Cipolli2018} or specifiying 
a location model \citep{Krzanowski1975, Daudin1999, Mbina2018}.

Another possible extension is to allow for $g>2$ response groups
which consequently involves a comparison of $O\{ (g/\ln(g))^g \}$ number of possible hypotheses \citep{GianCarlo1964}.
This increases the computational burden of our algorithm drastically and therefore would be pursued as future research.


By taking these into account, we believe that \texttt{VNPDA} has great potentials in 
analysing many high dimensional datasets when the normality assumption is questionable.
 
\section*{Funding information and conflict of interest}
This research was partially
supported by an Australian Postgraduate Award (Weichang Yu); 
and the Australian Research Council Discovery Project grant DP170100654 (John T. Ormerod).

\medskip 
The authors declare that they have no conflict of interest.
\appendix

\section{Derivation for algorithm in Table 2}

\label{Table1derivations}
We shall present the derivation in the case where we have a single new observation $(\vx_{n+1}, y_{n+1})$ 
and describe the generalisation to multiple new observations towards the end.
Following (\ref{prodComp1}), the updates for the parameter $\omega_j$ may be computed as
\begin{align} 
\label{exactomega}
\omega_j &= q(\gamma_j = 1), \nonumber \\[1ex]
&= \expit\left [ \bE_{-q(\gamma_j)} \left \{ \ln (1 + \vone^T \vgamma_{-j} ) - \ln (p^u + p - \vone^T \vgamma_{-j} - 1)   
+ \ln \mbox{BF}_j \right \} \right ].
\end{align}

\noindent 
The expression involves an expectation of the log Bayes factor
$$
\begin{array}{l} 
\ln \mbox{BF}_j 
= \sum_{\ell = 0}^{\infty} \sum_{\epsilon \in \text{\tt bin}(\ell)} 
\bigg[  \\ \quad
\phantom{+}
\ln\sB\Big\{ 
\alpha_{j, \epsilon0} + n_{j, \epsilon 0}^{(1)} + I(y_{n+1} = 1, x_{n+1,j} \in B_{j, \epsilon0}), \\ \quad
\qquad \quad \
\alpha_{j, \epsilon1} + n_{j, \epsilon 1}^{(1)} + I(y_{n+1} = 1, x_{n+1,j} \in B_{j, \epsilon1}) 
\Big\} 
\\ \quad
+ \ln\sB\Big\{ 
\alpha_{j, \epsilon0} + n_{j, \epsilon 0}^{(0)} + I(y_{n+1} = 0, x_{n+1,j} \in B_{j, \epsilon0}), \\ \quad
\qquad \quad \
\alpha_{j, \epsilon1} + n_{j, \epsilon 1}^{(0)} + I(y_{n+1} = 0, x_{n+1,j} \in B_{j, \epsilon1}) 
\Big\} \\ \quad
- \ln\sB\Big\{ 
\alpha_{j, \epsilon0} + n_{j, \epsilon 0} + I(x_{n+1,j} \in B_{j, \epsilon0}), \\ \quad
\qquad \quad \ 
\alpha_{j, \epsilon1} + n_{j, \epsilon 1} + I(x_{n+1,j} \in B_{j, \epsilon1}) 
\Big\} - \ln \sB(\alpha_{j, \epsilon0}, \, \alpha_{j, \epsilon1}) \bigg],
\end{array} 
$$

\noindent 
where {\tt bin}$(\ell)$ is the set of all binary representations of length $\ell$, $I( \cdot )$ is the indicator function, 
$\sB(a,b) = \Gamma(a) \Gamma(b) / \Gamma(a+b)$ is the Beta function, $n_{j, \epsilon}^{(k)}$ is
the number of group $k$ observations in the partition-subset $B_{j, \epsilon}$ and 
$n_{j, \epsilon} = n_{j, \epsilon}^{(1)} + n_{j, \epsilon}^{(0)}$.

Clearly, the expectation in equation $(\ref{exactomega})$ involves nonlinear functions of $\vgamma_{-j}$
and $y_{n+1}$. Therefore, we shall utilise some approximations to get around this.
We may use Taylor's expansion about $\vone^T \vomega_{-j}$ to approximate
$$
\begin{array}{rcl}
\bE_{-q_j} \ln (1 + \vone^T \vgamma_{-j} ) & \approx & \ln (1 + \vone^T \vomega_{-j} ),\\ [1ex]
\bE_{-q_j} \ln (p^u + p - \vone^T \vgamma_{-j} - 1)  & \approx & \ln (p^u + p - \vone^T \vomega_{-j} - 1),
\end{array} 
$$

\noindent 
and, for small $1/n$, a Stirling's approximation to approximate the Beta functions
\begin{align*}
&\ln \sB\Big \{ \alpha_{j, \epsilon0} 
+ n_{j, \epsilon 0}^{(1)} + I(y_{n+1} = 1, x_{n+1,j} \in B_{j, \epsilon0}),\\  &\hspace{2cm} \alpha_{j, \epsilon1} + n_{j, \epsilon 1}^{(1)} + I(y_{n+1} = 1, x_{n+1,j} \in B_{j, \epsilon1})\Big \} \\
&\quad \approx \ln \sB(\alpha_{j, \epsilon0} 
+ n_{j, \epsilon 0}^{(1)} , \, \alpha_{j, \epsilon1} + n_{j, \epsilon 1}^{(1)} ).
\end{align*}
Hence, the update equation for $\omega_j$ may be approximated as
\begin{align*}
\omega_j
&\approx \expit \left [ \ln (1 + \vone^T \vomega_{-j} ) - \ln (p^u + p - \vone^T \vomega_{-j} - 1)   
+ \ln \mbox{BF}_j \right]
\end{align*}
where
\begin{align*}
&\ln \mbox{BF}_j 
 \approx \sum_{\ell= 0}^{\infty} \sum_{\epsilon \in \text{\tt bin}(\ell)} \bigg \{ \ln \sB(\alpha_{j, \epsilon0} 
+ n_{j, \epsilon 0}^{(1)}, \, \alpha_{j, \epsilon1} + n_{j, \epsilon 1}^{(1)}) - \ln \sB(\alpha_{j, \epsilon0}, \, \alpha_{j, \epsilon1}) \\ &+ \ln \sB(\alpha_{j, \epsilon0} 
+ n_{j, \epsilon 0}^{(0)}, \, \alpha_{j, \epsilon1} + n_{j, \epsilon 1}^{(0)})
- \ln \sB(\alpha_{j, \epsilon0} + n_{j, \epsilon 0}, \, \alpha_{j, \epsilon1} + n_{j, \epsilon 1}) 
\bigg \}.
\end{align*}

\noindent 
The sum to infinity in the above expression is computationally tractable as
the subset counts decreases as we go further down the layers of
the tree. In particular, there exists a constant $M_j$ such that either $n_{j, \epsilon}^{(1)} = 0$ or $n_{j, \epsilon}^{(0)} = 0$
for all $\epsilon \in \bigcup_{\ell > M_j}$ {\tt bin}$(\ell)$ and $k \in \{0,1 \}$. Hence, we may rewrite the log Bayes factor as
\begin{align}
\label{approxlogBF}
&\ln \mbox{BF}_j = \sum_{\ell = 0}^{M_j} \sum_{\epsilon \in \text{\tt bin}(\ell)} \bigg \{ \ln \sB(\alpha_{j, \epsilon0} 
+ n_{j, \epsilon 0}^{(1)}, \, \alpha_{j, \epsilon1} + n_{j, \epsilon 1}^{(1)}) - \ln \sB(\alpha_{j, \epsilon0}, \, \alpha_{j, \epsilon1}) \nonumber \\ 
&+ \ln \sB(\alpha_{j, \epsilon0} 
+ n_{j, \epsilon 0}^{(0)}, \, \alpha_{j, \epsilon1} + n_{j, \epsilon 1}^{(0)})
- \ln \sB(\alpha_{j, \epsilon0} + n_{j, \epsilon 0}, \, \alpha_{j, \epsilon1} + n_{j, \epsilon 1}) 
 \bigg \}.
\end{align}

Similarly, we use (\ref{prodComp1}) to derive the update for $y_{n+1}$ as
\begin{align}
\label{classificatonRule}
\psi &= q(y_{n+1} = 1), \nonumber \\
&= \expit \bigg [ \bE_{-q(y_{n+1})} \bigg\{  \ln \bigg ( \frac{a_y + n_1 + I(y_{n+1} = 1)}{b_y + n_0 + I(y_{n+1} = 0)} \bigg )
 \nonumber \\
&\hspace{2cm} + \vgamma^T  \ln \big( \vpi^{(1)} \big)  - \vgamma^T \ln \big( \vpi^{(0)} \big) \bigg\} \bigg],
\end{align}

\noindent and if $n$ is large, then 
\begin{align} 
&\approx \expit \left[  \ln \bigg ( \frac{a_y + n_1}{b_y + n_0} \bigg )
+ \vomega^T  \ln \big( \vpi^{(1)} \big)
- \vomega^T \ln \big( \vpi^{(0)} \big) \right], 
\end{align}

\noindent 
where the number of observations in groups 1 and 0 are $n_1$ and $n_0$ respectively, the vector $\vpi^{(k)}$ of size $p$ is 
such that the j-th element is
\begin{align*}
\pi_{j}^{(k)} 
& = \exp\left[ 
\sum_{\ell = 0 }^{N_j} \ln \bP_k \big\{ B_{j, \epsilon_{j}(\ell)} \rightarrow B_{j, \epsilon_{j}(\ell+1)}  \big\} 
\right], 
\\
&= \exp\Bigg[ 
\sum_{\ell = 0}^{N_j} \bigg\{ \ln \Big(  
\alpha_{j, \epsilon_{j}(\ell+1)} + 
n_{j, \epsilon_{j}(\ell+1)}^{(k)} + 
I(y_{n+1} = k, x_j^* \in B_{\epsilon_j(\ell+1)}) 
\Big)
\\ 
& \hspace{12mm} 
- \ln \Big( 2\alpha_{j, \epsilon_{rj}(\ell+1)} + 
n_{j, \epsilon_{j}(\ell) }^{(k)} 
+ I(y_{n+1} = k, x_j^* \in B_{\epsilon_j(\ell)}) \Big) \bigg\}  \Bigg], 
\\
\text{and} \nonumber \\
&\approx \exp \left[ \sum_{\ell = 0 }^{N_j} \bigg\{ \ln \Big( \alpha_{j, \epsilon_{j}(\ell+1)} + n_{j, \epsilon_{j}(\ell+1)}^{(k)} \Big)
- \ln \Big( 2\alpha_{j, \epsilon_{j}(\ell+1)} + n_{j, \epsilon_{j}(\ell) }^{(k)} \Big) \bigg\}  \right],
\end{align*}

\noindent 
the $\ln$ prefix of a vector denotes element-wise $\ln$,
the binary representation $\epsilon_{j}(\ell)$ denotes the first $\ell$ branching directions taken by $x_{n+1,j}$
and the number $N_j$ is such that $n_{j, \epsilon_{j}(\ell) }^{(k)} = 0$ for all $\ell > N_j$ and $k \in \{0,1\}$.
Note that the approximation in equation (\ref{classificatonRule})
follows from Taylor's approximation.

We may extend the classification rule in equation (\ref{classificatonRule}) to $m>1$ new observations by simply
replacing each $\vpi^{(k)}$ with $\vpi_r^{(k)}$, where the j-th element of $\vpi_r^{(k)}$ is
\begin{align*}
\pi_{rj}^{(k)} &= \exp \left [ \sum_{ \ell = 0 }^{N_j} \ln \bP_k \big \{ B_{j, \epsilon_{rj}(\ell)} \rightarrow B_{j, \epsilon_{rj}(\ell+1)} \big \} \right ], \\
&= \exp\Big[ 
\sum_{\ell = 0 }^{N_j} \bigg\{ \ln \Big( 
\alpha_{j, \epsilon_{rj}(\ell+1)} 
+ n_{j, \epsilon_{rj}(\ell+1)}^{(k)} 
+ I(y_{n+r} = k, x_{n+r,j} \in B_{\epsilon_{rj}(\ell+1)}) 
\Big)
\\ &\hspace{12mm} 
- \ln\Big( 
2\alpha_{j, \epsilon_{rj}(\ell+1)} 
+ n_{j, \epsilon_{rj}(\ell) }^{(k)} 
+ I(y_{n+r} = k, x_{n+r,j} \in B_{\epsilon_{rj}(\ell)}) 
\Big) \bigg\} \Bigg ], 
\\
& \approx \exp\left[ \sum_{\ell = 0 }^{N_j} \bigg\{  
\ln\Big( 
\alpha_{j, \epsilon_{rj}(\ell+1)} 
+ n_{j, \epsilon_{rj}(\ell+1)}^{(k)} 
\Big)
- \ln\Big( 2\alpha_{j, \epsilon_{rj}(\ell+1)} + n_{j, \epsilon_{rj}(\ell) }^{(k)} \Big) \bigg\} \right],
\end{align*}

\noindent 
the number $N_j$ is such that $n_{j, \epsilon_{rj}(\ell) }^{(k)} = 0$ for all $\ell > N_j$ and $1 \le r \le m$ and 
the rest of the notations used have been explained in Section \ref{Inference}.

\bigskip 
\textbf{Remark:} In our numerical examples, we truncate the P{\'o}lya tree priors at layers $N_j = M_j = \ln_2 (n)$ for all $1 \le j \le p$. This allows us to account for the details at 
higher resolution of the group-conditional distributions as $n$ increases \citep{Hanson2002}.

\section*{References} 

\bibliographystyle{elsarticle-harv}

{\small 
\bibliography{VNPDAReferences}
}


%
%
%
\end{document}